\documentclass[aps,twocolumn,showpacs,preprintnumbers,amsmath,amssymb]{revtex4}
\usepackage{epsfig}
\usepackage{graphicx}
\usepackage{dcolumn}
\usepackage{bm}
\usepackage{amsmath}
\usepackage{amssymb}
\begin{document}
\title{The influence of the spin-dependent phases of tunneling electrons on
the conductance of a point  ferromagnet/isolator/d-wave
superconductor contact}
\author{B.P. Vodopyanov}
\affiliation{Kazan Physical-Technical Institute, Russian Academy
of Sciences, 10/7 Sibirsky Tract, 420029 Kazan, Russia}
\date{\today}
\begin{abstract}
The influence of the spin-dependent phase shifts (SDPS) associated to
the electronic reflection and transmission amplitudes acquired by
electrons upon scattering on the potential barrier on the Andreev
reflection probability of electron and hole excitations for a
ferromagnet/isolator/d-wave superconductor (FIS) contact and the
charge conductance of the FIS contact is studied. Various
superconductor orientations are considered. It is found that SDPS can
suppress the zero-potential peak and restore finite-potential peaks
in the charge conductance of the F/I/d-wave superconductor contact
for the (110) orientation of the d-wave superconductor and, on the
contrary, can restore the zero-potential peak and suppress
finite-potential peaks for the $\{100\}$ orientation of the d-wave
superconductor.
\end{abstract}

\pacs{74.50.+r, 74.80.-g, 74.80.Dm, 74.80.Fp, 75.30.Et}

\maketitle
\section{\qquad Introduction}
The oscillating character of the spatial dependence of the anomalous
Green function  (GF) in a ferromagnet  in various hybrid structures
containing the ferromagnet/superconductor (F/S) interface with a
singlet order parameter is due to the presence  of electron spin
subbands with different values of Fermi momenta  $p_{\,\alpha }$ in a
ferromagnetic metal (F) \cite{BuzdinJTPL, Demler, Kupriyanov,
BuzdinRMP} ($\,\alpha $ ($\,\alpha = \uparrow, \downarrow $) \,\,is
the spin index, which denotes the projection of the electron spin on
the direction of the magnetic moment of a ferromagnet). Such
manifestation of the proximity effect is the basis for creation of
the $\pi$-Josephson junction \cite{Ryazanov1}, various spin-valve
schemes \cite{BeasleyAPL, Johnson, Tagirov, BuzdinEl, Nazarov}, being
the main elements of promising superconducting electronics
\cite{BeasleyIEEE, Ustinov, Ioffe}.

The suppression of the Andreev reflection \cite{Andreev} in point F/S
contacts \cite{Beenakker} due to the decrease of the number of
conducting channels is another consequence of the presence of spin
subbands in a ferromagnetic metal. This fact is used to determine the
spin polarization of ferromagnetic materials \cite{ Buhrman1, Soulen,
Perez, Krivoruchko, Buhrman2}, to study the order parameter symmetry
of high-temperature superconductors \cite{Vasko, Dong, Chen, Luo} and
to control the spin-polarized currents \cite{Jedema, Urech,
Appelbaum}.

Recently an attention was paid to one more property of hybrid F/S
structures: the influence of spin-dependent phase shifts
$\theta_{\,\alpha}^{\,d}$ and $\theta_{\,\alpha}^{\,r}$ (SDPS)
associated to the electronic reflection and transmission amplitudes
$r_{\,\alpha}$ and $d_{\,\alpha} $ on the contact on thermodynamic
\cite{Sauls1988} and transport \cite{Fogelstrom, Barash, Sauls2004}
characteristics of hybrid structures with a spin-active interface:
\begin{equation*}
 d_{\,\alpha}\,=\,\sqrt{D_{\,\alpha}}\,
\,exp\,(i\,\theta_{\,\alpha}^{\,d}));\quad r_{\,\alpha}\,=\,
\sqrt{R_{\,\alpha}}\,\,exp\,(i\,\theta_{\,\alpha}^{\,r}).
\end{equation*}
Here $D_{\,\alpha}$ and $R_{\,\alpha}\,=\,1-D_{\,\alpha}$ are
transmission and reflection coefficients, respectively. Let us note,
that parameters  $d _ {\, \alpha} $ and $r _ {\, \alpha} $ are almost
insensitive to the appearance of  a  superconducting state and the
corresponding changes  will not be considered in the present paper.

It has been found that the difference of SDPS due to the difference of
potential barriers for electrons with different spin projections
$\,\alpha$ results in the appearance of a $\pi$ state in the  S/FI/S
junction (FI is a ferromagnetic isolator) without taking into account
the proximity effect \cite{Fogelstrom, Barash}.

The presence of SDPS also leads to the formation of spin-dependent
Andreev bound states $\varepsilon_{\,\alpha}^{\,b} ={\rm
{sign}}(p_{x,\,\alpha }-p_{x,\,-\,\alpha })\,\, \Delta
\cos((p_{x,\,\alpha }-p_{x,\,-\,\alpha })/2)$  in the N/F/s-wave
superconductor contacts (N is a normal metal; $ p_{x,\,\alpha }$  is
the projection of the Fermi momentum in a ferromagnet on the $x$
axis, being perpendicular to the contact plane) \cite{Sauls2004}. In
the tunneling limit these states appear as the resonance peaks below
the gap in the dependence of the ballistic charge conductance on the
applied potential $V$.

The influence of SDPS on the charge conductance  of a single-channel
quantum point contact of a F/s-wave superconductor and that of a
multichannel ballistic contact of a F/I/s-wave superconductor (I is
isolator) was studied in Refs. \cite{Cottet, Vod}, correspondingly.
In Ref. \cite{Cottet} it has been found that for a weakly transparent
contact, SDPS induces subgap resonances in the charge conductance of
the quantum point contact. For high transparencies, these resonances
are smoothed, but the shape of the signals remains extremely
sensitive to SDPS.  In Ref. \cite{Vod} it has been found that when F
is strongly polarized, the peak in the conductance of the F/I/s-wave
superconductor contact can be restored at a zero potential.

Such strong influence of SPDS on transport properties of hybrid
structures with ferromagnetic elements allows one to suppose that
they may be successfully used in experiments on Andreev spectroscopy
of ferromagnets, superconductors and in various applications in the
field of nanospintronics.

This paper is devoted to a theoretical study of the SDPS influence on the
Andreev reflection and charge conductance of a point
F/I/d-wave superconductor contact.

Superconductors with the $d$-wave symmetry (the $ d_{x^2 - y^2}$
symmetry of the order parameter is considered) have an internal,
momentum-dependent phase, which strongly influences the transport
properties of contacts between them and other materials. In Ref.
\cite{Hu} it was shown that when the angle $\gamma $ between the $a$
axis of a superconducting crystal and the normal to the surface of
the high-ohm  interface is $ \pi /4 $ (the $\{110\}$ orientation of
the d-wave superconductor), then a bound state is formed on the Fermi
level near the high-ohm interface. This zero-energy bound state
resulting from the repeated Andreev reflections \cite{Harlingen,
Kashiwaya2} causes a sharp peak at a zero potential in the dependence
of the charge conductance of the N/I/d-wave superconductor on the
applied potential \cite{Kashiwaya}.

The first theoretical study of spin-polarized tunneling spectroscopy
of F/I/d-wave superconductor junctions was performed in Refs.
\cite{Zhu, Beasley, Zutic}. It has been found that the subgap charge
conductance behavior is qualitatively different from a nonmagnetic
case. In particular, it has been found that for the $\{110\}$
orientation of the d-wave superconductor the zero-potential peak in
the charge conductance is suppressed by the exchange interaction due
to the suppression of Andreev reflections and that it splits into two
peaks under the influence of the exchange interaction in the
insulator.

The influence of SDPS $\theta_{\,\alpha}^{\,d}$ and
$\theta_{\,\alpha}^{\,r}$ associated to the electronic reflection and
transmission amplitudes $r$ and $d$ on the contact on the charge
conductance of the F/I/d-wave superconductor contact in Refs.
\cite{Zhu, Beasley, Zutic} is not studied.

The main result of this paper is that SDPS can suppress the
zero-potential peak and restore finite-potential peaks in the charge
conductance of the F/I/d-wave superconductor contact for the
$\{110\}$ orientation of the d-wave superconductor and, on the
contrary, can restore the zero-potential peak and suppress
finite-potential peaks for the $\{100\}$ orientation of the d-wave
superconductor. This takes place because due to the interference of
one part of trajectories of electron-like and hole-like
quasiparticles reflected by the pair potential and the interface,
spin-dependent bound states are formed near the Fermi level, whereas
due to the interference of the other part of trajectories
spin-dependent bound states are formed in the vicinity of edges of
the energy gap. Spin-dependent amplitudes of the Andreev reflection
probability and energy levels of spin-dependent Andreev bound states
are also found.

This work illustrates that the study of the influence of SDPS on the
charge conductance of the point F/I/d-wave superconductor contact can
provide an interesting insight in the spin-dependent transport.

A theoretical possibility to study the influence of SDPS associated
to the electronic reflection and transmission amplitudes acquired by
electrons upon scattering on the potential barrier on the $I\,-\,V$
characteristics of superconducting weak links  with ferromagnetic
elements appeared after the boundary conditions (BCs) for the
quasiclassical GF were obtained. In Ref. \cite{Sauls1}, BCs for the
quasiclassical GF for two metals in contact via a magnetically active
interface in terms of an interface scattering matrix were derived. In
Ref. \cite{Fogelstrom}, BCs for the retarded and advanced
quasiclassical GFs were obtained in terms of Riccati amplitudes
\cite{Eschrig, Shelankov}. In Ref. \cite{Sauls2004}, BCs in terms of
Riccati amplitudes were obtained for the nonequilibrium
quasiclassical GF. In Ref. \cite{Vod2003}, quasiclassical equations
of superconductivity for metals with a spin-split conduction band
were derived and BCs for the temperature quasiclassical GF for the
F/S interface were obtained. The model interface was the same as in
Refs. \cite{Sauls1, Zaitsev}.

In this paper, calculations are carried out using quasiclassical GFs
and the relevant BCs obtained in Ref. \cite{Vod2003}.
\section{Finding differential conductance of a point FIS contact}
\subsection{\label{sec:level1} The general expression for differential
conductance of a point contact through quasiclassical GF} In hybrid
F/S structures the Andreev reflection is modified. The reflected hole
has some parameters (for example, the velocity modulus and the phase
shift) different from those of the incident electron because it moves
in a subband with an opposite spin. Such spin-discriminating
processes due to the exchange field in a ferromagnet lead to the
formation of spin-dependent Andreev bound states inside the gap
\,\cite{Barash, Fogelstrom}.  As a result, the spectral density of
the charge conductance $ G_{FIS} $ of the FIS contact at a zero
potential is no longer  a symmetrical function of energy
$\varepsilon$. The condition of the time reversal invariance has the
form $G_{FIS}(\varepsilon,\,\alpha) $\,=\,$
G_{FIS}(-\,\varepsilon,\,-\,\alpha) $.  The generalization of the
charge conductance expression \cite{Tinkham} for this case  results
in the following formula for $ G_{FIS}(V)$ \cite{Vod}:
\begin{gather}
G_{FIS}(V)=\frac{e^{\,2}A}{32\pi^{\,2}\,T}\sum_\alpha\,\rm{Tr}
\left[ \int\frac{d{\bf p}_\|}{\,(2\,\pi)^{\,2}}
\int\limits_{-\infty}^{\infty} d\varepsilon\,\times \right. \notag
\\
\left.
\frac{1}{\coth^2(\frac{\varepsilon-eV\,\hat{\tau}_z}{2\,T})}\,\,
[1-\hat{g}_{\,s}^{\,A}\, \tau_z \,\hat{g}_{\,s}^{\,R}
\,\hat{\tau}_{\,z} -\hat{g}_{\,a}^{\,A}\,\hat{\tau}_z
\,\hat{g}_{\,a}^{\,R}\,\hat{\tau}_{\,z}\right. \notag
\\
\left. +\hat{\Upsilon}_{\,s}^{\,A}\hat{\tau}_z
\hat{\Upsilon}_{\,s}^{\,R}\,\hat{\tau}_{\,z}-\hat{\Upsilon}_{\,a}^{\,A}
\hat{\tau}_z \hat{\Upsilon}_{\,a}^{\,R} \hat{\tau}_{\,z}]\right].
\label{eq:1}
\end{gather}
In Eq. (\ref{eq:1}), $V$ is the potential;\,$A$\, is the contact
area; $e$ is the electron charge; $T$ is the temperature;
$\hat{\tau}_z$ is the Pauli matrix; \, $p_\|$ is the momentum in the
contact plane;\, ($\hat{g}_{\,s}$, $\hat{\Upsilon}_{\,s}$) and
\,($\hat{g}_{\,a} $, $\hat{\Upsilon}_{\,a}$) are quasiclassical
retarded (R) and advanced (A) GFs symmetric (s) and antisymmetric (a)
\,\cite{Vod} with respect to the projection of the momentum \,$
{\bf\hat{p}}$ \, on the Fermi surface on the $x$ axis, being
perpendicular to the contact plane, composed according to the rule
$\hat{T}_{\,s(a)}\,=\,1/2\,[\hat{T}( p_{\,x})\,\pm \,\hat{T}(-\,
p_{\,x})]$.

Besides the matrix quasiclassical GF $\hat{g}$, which equation is
analogous to that derived in Ref. \cite{Larkin}, equation
(\ref{eq:1}) includes the matrix GF $\hat{\Upsilon}$, describing the
interference of waves incoming to the interface and outgoing from it.
The function relation with the matrix one-particle temperature GF and
equations, which the function obeys, are presented in appendix.
Calculations in Eq.(\ref{eq:1}) are to be carried out on the boundary
of any contacting metal.
\subsection{\label{sec:level2} Finding quasiclassical GF}
Let us assume that the barrier with the width $d$ is located in the
region $ -\,d/2\,<x<\,d/2$, the superconductor occupies the region
$x>d/2$, and the ferromagnet occupies the region $x<-\,d/2$. To find
GFs for each metal, one has to solve quasiclassical equations of
superconductivity for metals with a spin-split conductivity band
simultaneously with their BCs derived  in Ref. \cite{Vod}:
\begin{multline}\label{eq:2}
{\rm{sign}}(\hat{p}_{\,x})\frac{\partial}{\partial{\,x}}\,\hat{g}+\frac{1}{2}\,
{\bf
v_\|}\frac{\partial}{\partial{\bf\rho}}(\hat{v}_{\,x}^{-1}\hat{g}+\hat{g}\,\hat{v}_{\,x}^{-1})
+[\hat{K},\,\hat{g}]_- = 0, \\
{\rm
{sign}}(\hat{p}_{x})\frac{\partial}{\partial{x}}\hat{\Upsilon}+\frac{1}{2}
{\bf
v_\|}\frac{\partial}{\partial{\bf\rho}}(\hat{v}_{x}^{-1}\hat{\Upsilon}-
\hat{\Upsilon}\hat{v}_{x}^{-1})+[\hat{K},\hat{\Upsilon}]_+ = 0,\\
\hat{K}=\,-\,i\hat{v}_{\,x}^{-\frac{1}{2}}(i\varepsilon_n
\hat{\tau}_z+
\hat{\Delta}-\hat{\Sigma})\hat{v}_{\,x}^{-\frac{1}{2}} -
i(\hat{p}_{\,x}-\hat{\tau}_x\hat{p}_{\,x}\hat{\tau}_x)/2, \\
\hat{\Delta}\,\equiv \,\hat{\Delta}(x,\,{\bf p}),\qquad
\shoveleft{[a,\,b]_\pm = ab \pm ba.} \qquad
\end{multline}
In this section, $\varepsilon_n = (2n + 1)\pi T $  is the Matsubara
frequency; $\hat{\tau}_{\,x}$ and $\hat{\tau}_{\,z}$ are the Pauli
matrices; $\rho=(x,y)$ are coordinates in the contact plane;  $\hat{\Sigma}$
is the self-energy part; $\hat{g}$ are matrix temperature GFs:
$$ \hat{g}=
  \begin{pmatrix}
    g_{\,\alpha,\,\alpha} & f_{\,\alpha,\,-\alpha} \\
    f_{-\,\alpha,\,\alpha}^{\,+}&\,-\, g_{-\,\alpha,\,-\alpha}
  \end{pmatrix}, \,\,
 \hat{g}=\left\{\begin{split}\hat{g}_>\qquad  \hat{p}_{\,x}>0,\\
\hat{g}_< \qquad  \hat{p}_{\,x}<0
\end{split}\right. .$$
Moreover,
$$
 \hat{\Delta} = \begin{pmatrix}
   0& \Delta (x,\,{\bf p})\\
    -\Delta^*(x,\,{\bf p}) &0
  \end{pmatrix},
  \;\hat{p}_{x}=
  \begin{pmatrix}
   p_{x,\,\alpha }& 0 \\
   0 & p_{x,\,{-\alpha}}
  \end{pmatrix},$$
where $\Delta (x,\,{\bf p})$ is the order parameter;\, $
p_{x,\,\alpha }$ and $ p_{\|}$ are projections of the momentum on the
Fermi surface on the $x$ axis and the contact plane, respectively;
 \,$\hat{v}_{\,x}$ =\,$ \hat{p}_{\,x}/m $ and
$\hat{v}_{\|}=p_{\|}/m $.

BCs for a specular reflection of electrons from the boundary
 $ p_\parallel $ = $p_{\downarrow}\sin\vartheta_\downarrow
$ =
 $p_{\uparrow}\sin\vartheta_\uparrow =
p_S\sin\vartheta_S $  have the form  \cite{Vod}:
\begin{gather}
(\hat{\tilde{g}}_a^S)_d = (\hat{\tilde{g}}_a^F)_d,\;\;
(\hat{\tilde{\Upsilon}}_a^S)_d =
(\hat{\tilde{\Upsilon}}_a^F)_d,\nonumber
\\
(\sqrt{\hat{R}_\alpha}-\sqrt{\hat{R}_{-\alpha}})(\hat{\tilde{\Upsilon}}_a^+)_n\nonumber
= \alpha_3(\hat{\tilde{g}}_a^-)_n,
\\
(\sqrt{\hat{R}_\alpha}-\sqrt{\hat{R}_{-\alpha}})(\hat{\tilde{\Upsilon}}_a^-)_n
= \alpha_4(\hat{\tilde{g}}_a^+)_n,\label{eq:3}
\\
-\hat{\tilde{\Upsilon}}_s^- =
\sqrt{\hat{R}_\alpha}(\hat{\tilde{g}}_s^+)_d+\alpha_1(\hat{\tilde{g}}_s^+)_n,\nonumber
\\
-\hat{\tilde{\Upsilon}}_s^+ =
(\hat{R}_\alpha)^{-\frac{1}{2}}(\hat{\tilde{g}}_s^-)_d+\alpha_2(\hat{\tilde{g}}_s^-)_n,\nonumber
\end{gather}
where\; $\hat{\tilde{g}}_{a(s)}^{\pm}
 = 1/2\,[\,{\hat{\tilde{g}}}_{a(s)}^S \pm {\hat{\tilde{g}}}_{a(s)}^F\,]$. \;
Functions $\hat{\tilde{\Upsilon}}_{a(s)}^{\pm}$ are determined
analogously. In Eq.  (\ref{eq:3}) and below, the indices $ d $ and $
n $ denote the diagonal and the nondiagonal parts of the matrix $(
\hat{T}_{d(n)} = 1/2\,[ \, \hat{T} \pm \tau_z\hat{T}\tau_z \,])$,
respectively. GFs \,\,$\hat{\tilde{g}}$ \, are connected with GFs, \,
being the solutions of Eq. (\ref{eq:2}), by the following
relationships \cite{Vod}:
\begin{gather}
(\hat{\tilde{g}}_s^S)_n\,=\,(\hat{g}_s^S)_n\,\,
\cos(\,\theta_{\,\alpha}) +i\hat{\tau}_z \,(\hat{g}_a^S)_n
\,\,\sin(\,\theta_{\,\alpha})\nonumber
\\
(\hat{\tilde{g}}_a^S)_n\,=\,(\hat{g}_a^S)_n\,\,\cos(\,\theta_{\,\alpha})
+i\hat{\tau}_z \,(\hat{g}_s^S)_n\,\, \sin(\,\theta_{\,\alpha})\nonumber\\
(\hat{\tilde{g}}_s^F)_n\,=\,(\hat{g}_s^F)_n\,\,\cos(\beta_{\,\alpha}^{\,r})
+i\hat{\tau}_z (\hat{g}_a^F)_n\,\, \sin(\beta_{\,\alpha}^{\,r}) \label{eq:4}\\
(\hat{\tilde{g}}_a^F)_n\,=\,(\hat{g}_a^F)_n\,\,
\cos(\beta_{\,\alpha}^{\,r}) +i\hat{\tau}_z (\hat{g}_s^F)_n\,\,
\sin(\beta_{\,\alpha}^{\,r})\nonumber
\\
(\hat{\tilde{\Upsilon}}^{F})_n\,=\,(\hat{\Upsilon}^{F})_n\,e^{\,i\,{\rm
{sign}}\,(\hat{p}_{\,x})(\theta_{\alpha}^{\,r}+
\theta_{\,-\,\alpha}^{\,r})/2}\nonumber
\\
\theta_{\,\alpha}=\frac{\theta_{\alpha}^{\,r}-
\theta_{-\alpha}^{\,r}}{2}-
(\theta_{\,\alpha}^{\,d}-\theta_{-\,\alpha}^{\,d});\quad \,\,
\beta_{\,\alpha}^{\,r}=\frac{\theta_{\alpha}^{\,r}-
\theta_{-\alpha}^{\,r}}{2}.\nonumber
\end{gather}
The diagonal parts of matrices $\hat{\tilde{g}}$ are equal to the
corresponding matrices $\hat{g}$. An explicit form of other functions
$\hat {\tilde{\Upsilon}}$ is not needed. These functions are found
from BCs. Coefficients $ \alpha_i$ are:
\begin{gather*}
\alpha_{1(2)} = \frac{1+\sqrt{{R_{\,\uparrow}}R_{\,\downarrow}}
\mp \sqrt{{D_{\,\uparrow}}D_{\,\downarrow}}}{\sqrt{R_{\,\uparrow}}
+ \sqrt{R_{\,\downarrow}}},\;\;
\\
\alpha_{3(4)} = 1-\sqrt{{R_{\,\uparrow}}R_{\,\downarrow}}\, \pm
\sqrt{{D_{\,\uparrow}}D_{\,\downarrow}}\;).\;
\end{gather*}
When solving Eqs. (\ref{eq:2}), let us assume that the order
parameter is  a step function, being zero in the ferromagnet and
finite in the superconductor. Then for S metal the solution is as
follows:
\begin{gather}\label{eq:5}
\hat{g}(x,{\bf p})= e^{\,-\,{\rm
{sign}}(\hat{p}_{\,x})\hat{K}(x-\frac{d}{2})}\hat{C}({\bf
p})e^{\,{\rm {sign}}(\hat{p}_{\,x})\hat{K}(x-\frac{d}{2})} +
\hat{C}_{\,0}({\bf p})\nonumber
\\
\hat{\Upsilon}(x,{\bf p}) = e^{\,-\,{\rm
{sign}}(\hat{p}_{\,x})\hat{K}(x-\frac{d}{2})}\hat{\Upsilon}e^{\,-{\rm
{sign}}(\hat{p}_{\,x})\hat{K}(x-\frac{d}{2})}\nonumber\\
\hat{\Upsilon}= \hat{\Upsilon}(\,x=0,{\bf p}).\label{eq:5}
\end{gather}
Matrices $ \hat{C}_{\,0}({\bf p}) $ are the values of  GFs $ \hat{g}
$ far from the F/S boundary:
\begin{gather}
\hat{C}_{\,0}^{\,S}({\bf p})\,=\,\begin{pmatrix}
  g & f \\
 f^{\,+}& -\,g
\end{pmatrix} = \frac{\begin{pmatrix}
  \varepsilon_n & -i\Delta ({\bf p}) \\
  i\Delta^*({\bf p}) & -\varepsilon_n
\end{pmatrix}}{\sqrt{\varepsilon_n^2
+ |\Delta({\bf p})|^2}}\nonumber
\\
\Delta({\bf
p})\,=\,\Delta_{\,d}(T)\,\cos(2\,\vartheta_{\,S}\,-\,2\,\gamma).\label{eq:6}
\end{gather}
In Eq. (\ref{eq:6})\,\,$\Delta_{\,d}(T)$ is the maximum value of the
order parameter at temperature $T$;\, $\vartheta_{\,S}$\, is the
angle between the electron momentum in the superconductor and the $x$
axis, being perpendicular to the contact plane, and  $ \gamma $ is the angle
between the crystal \,\,$a$ \,\,axis of the  $d$-wave superconductor
and the $x$ axis.

For F metal the solution has the same form as Eq. (\ref{eq:5}) except
for changing the exponent argument from $(x-d/2)$ to $(x+d/2)$;\quad
$\hat{C}_{\,0}^{\,F} =
\hat{\tau}_z\,\,\varepsilon_n/|\varepsilon_n|.$

GFs $ \hat{g}^{\,S} $ in Eq. (\ref{eq:5}) have to tend to
$\hat{C}_{\,0}^{\,S} $ at $ x \rightarrow +\,\infty $ and GFs
$\hat{g}^{\,F}$ to  $ \hat{C}_{\,0}^{\,F}({\bf p})$ at $ x\rightarrow
-\,\,\infty $. By matrix multiplication in Eq. (\ref{eq:5}) and in
corresponding equation for $ \hat{g}^{\,F} $, we find that for the
above to hold it is necessary that at $x=+\,d/2$ and at $x=-\,d/2$
the relationships
\begin{gather}
\hat{C}_{\,0}^{\,S}({\bf p}) \hat{C}^{\,S}({\bf
p})\,=\hat{C}^{\,S}({\bf p})\,\hat{C}_{\,0}^{\,S}({\bf p})={\rm
{sign}}(\hat{p}_{\,x}) \,\hat{C}^{\,S}({\bf p})\label{eq:7}\\
\hat{C}_{\,0}^{\,F}({\bf p}) \hat{C}^{\,F}({\bf p})\,=
 -\hat{C}^{\,F}({\bf p})\,\hat{C}_{\,0}^{\,F}({\bf p})=
 \,-\, {\rm {sign}}(\hat{p}_{\,x}) \hat{C}^{\,F}({\bf p})\qquad \nonumber
\end{gather}
are fulfilled respectively. It follows from these relationships that
\begin{gather}
\hat{g}_{\,s}^{\,S}\, = \,\hat{X}\,\hat{C}_{\,a}^{\,S}\,+\,\hat{X},
\qquad \hat{g}_{\,s}^{\,F}\, = \,\hat{C}_{\,0}^{\,F}\,-\,
\hat{C}_{\,0}^{\,F}\,\hat{C}_{\,a}^{\,F}\nonumber
\\
\hat{g}_{\,a}^{\,S}\,=\,\hat{C}_{\,a}^{\,S}\,+\hat{C}_{\,0,\,a}^{\,S},\,\qquad\quad
\hat{g}_{\,a}^{\,F}\, \equiv \hat{C}_{\,a}^{\,F},\qquad \quad
\label{eq:8}
\end{gather}
where \begin{equation*}
 \hat{X}=
(1+\hat{C}_{\,0,\,a}^{\,S})(\hat{C}_{\,0,\,s}^{\,S})^{-\,1},\qquad
\hat{X}=\hat{\tau}_z\,(X)_d+(\hat{X})_n.
\end{equation*}
In Eq. (\ref{eq:8}) $\hat{C}_{\,0,\,s(a)}^{\,S}$\,\, are symmetric
and antisymmetric combinations of  the matrix
$\hat{C}_{\,0}^{\,S}({\bf p})$ with respect to the projection of the
Fermi momentum on the $x$ axis:\,\,\,\,
$\hat{C}_{\,0,\,s(a)}^{\,S}\,=\,1/2\,[\hat{C}_{\,0}^{\,S}(
p_{\,x})\,\pm \,\hat{C}_{\,0}^{\,S}(-\, p_{\,x})]$,\,
\,$\hat{X}\,=\,\hat{1}$. \,\, Matrices $\hat{\Upsilon}^{S(F)}$
satisfy the relationships:
\begin{multline}\label{eq:9}
\hat{C}_{\,0}^{\,F}({\bf p}) \hat{\Upsilon}^{F}({\bf p}) =
\hat{\Upsilon}^{F}({\bf p}) \hat{C}_{\,0}^{\,F}({\bf p}) =
\,-\,{\rm{sign}}(\hat{p}_{\,x})({\bf p}) \hat{\Upsilon}^{F}({\bf p})\\
\hat{C}_{\,0}^{\,S}({\bf p})\hat{\Upsilon}^{S}({\bf p})=
\hat{\Upsilon}^{S}({\bf p})\hat{C}_{\,0}^{\,S}({\bf p})= {\rm
{sign}}(\hat{p}_{\,x})({\bf p})\hat{\Upsilon}^{S}({\bf p}),
\end{multline}
being the condition for the functions $\hat{\Upsilon}^{F}(x,{\bf p})$
and  $\hat{\Upsilon}^{S}(x,{\bf p})$ to tend to zero when $x$ tends
to  $-\,\infty$ and $+\,\infty$, respectively. It follows from Eq.
(\ref{eq:2}) that the function $(\hat{\Upsilon}(x)^{F})_n =
const\,=\,0$, because for a ferromagnet
$[\hat{K},(\hat{\Upsilon})_n]_+\,=\,0$ and $\hat{\Upsilon}^{F}(x,{\bf
p})$ has to tend to zero when $x$ tends to $-\,\infty$. Then from the
BCs (\ref{eq:3}) and relationships (\ref{eq:4}) it follows that:
\begin{equation}\label{eq:10}
 \alpha_3(\hat{\tilde{g}}_a^-)_n = \alpha_4
 (\hat{\tilde{g}}_a^+)_n,\qquad
\alpha_1 (\hat{\tilde{g}}_s^+)_n = \alpha_2
(\hat{\tilde{g}}_s^-)_n.
\end{equation}
From the first equality in Eq. (\ref{eq:10}) we find the relation
between functions $(\hat{\tilde{g}}_{\,a}^{\,F})_n$ and
$(\hat{\tilde{g}}_{\,a}^{\,S})_n$:
\begin{equation*}(\hat{\tilde{g}}_{\,a}^{\,F})_n\,=\,
\frac{\sqrt{D_{\uparrow}D_{\downarrow}}}
{1-\sqrt{{R_{\,\uparrow}}R_{\,\downarrow}}}\,(\hat{\tilde{g}}_{\,a}^{\,S})_n.
\end{equation*}
By substituting this relation into the second equality in Eq.
(\ref{eq:10}) and using the relations (\ref{eq:4}) and
(\ref{eq:8}) we find $(\hat{\tilde{g}}_{\,a}^{\,F})_n$:
\begin{gather}
\hat{\tilde{g}}_a^F\,=\,\hat{g}_a^F\,e^{\,-i\,\beta_{\,\alpha}^{\,r}\,{\rm
{sign}}(\varepsilon_n)}=
-\,\frac{\sqrt{D_{\uparrow}D_{\downarrow}}\,\hat{\tau}_z\,(\hat{X})_n\,
  }{Z} \nonumber\\
Z=(1-\sqrt{R_\uparrow R_\downarrow})\,[X_d  \,
\cos(\theta_{\,\alpha})+i\, \sin(\theta_{\,\alpha})] \label{eq:11}\\
+\,(1+\sqrt{R_\uparrow
R_\downarrow})\,{\rm{sign}}(\varepsilon_n)\,
[\cos(\theta_{\,\alpha})\,+i\,X_d \,
\sin(\theta_{\,\alpha})].\nonumber
\end{gather}
Knowing $(\hat{\tilde{g}}_{\,a}^{\,F})_n$, from Eqs. (\ref{eq:3}) and
(\ref{eq:4}) we find functions $\hat{g}_{\,s}^{\,F},
\hat{\Upsilon}_{\,s}^{\,F}$ and $\hat{\Upsilon}_{\,a}^{\,F}$
necessary for calculation of the conductivity in the Eq. (\ref{eq:1})
and calculate the conductance at the ferromagnet side.
\subsection{\label{sec:level3} Differential conductance of a point FIS contact}
After carrying out the analytical continuation in functions
$(\hat{\tilde{g}}_{\,a}^{\,F})_n, \hat{g}_{\,s}^{\,F},
\hat{\Upsilon}_{\,s}^{\,F}, \hat{\Upsilon}_{\,a}^{\,F}$ (substitution
$i\,\varepsilon_n $\, for\, $ \varepsilon \pm \delta \,$ for retarded
and advanced GFs, respectively), we obtain an expression for the
charge conductance $\sigma_{\,F/S}(V)$. For angles $\gamma = 0$ and
$\gamma = \pi/4$ \,\, $\sigma_{\,F/S}(V)$ \,is as follows:
\begin{gather}
\sigma_{\,F/S}(V)=\frac{\,e^{\,2}\,A}{\pi}\int\frac{d{\bf
p}_{\|}\,}{\,(2\,\pi)^{\,2}}\left\{\,\,
\int\limits_{|\Delta(\vartheta_S)|}^\infty\frac{d\,\varepsilon}{2\,T}
\left[ \frac{1}{\cosh^{\,2}(\frac{\varepsilon+eV}{2\,T})}\,\right.
\right.\notag
\\
\left.  +\frac{1}{\cosh^{\,2}(\frac{\varepsilon-eV}{2\,T})}
\right]\frac{\varepsilon\,\xi^{\,R}(D_{\uparrow}+D_{\downarrow})+
\varepsilon\,(\varepsilon-\xi^{\,R})D_{\uparrow}\,D_{\downarrow}}
{Z_\Uparrow}\,\,+\label{eq:12}
\\
\int\limits_0^{|\Delta(\vartheta_S)|}\frac{d\varepsilon}{2T}
\left[
\frac{D_{\uparrow}D_{\downarrow}}{\cosh^{2}(\frac{\varepsilon+eV}{2T})}+
\frac{D_{\uparrow}D_{\downarrow}}{\cosh^{2}(\frac{\varepsilon-eV}{2T})}
\right] \left.\frac{
|\Delta(\vartheta_S)|^{2}}{Z_\Downarrow}\right\}.\notag
\end{gather}
For $\gamma = 0$:
\begin{gather}
\Delta(\vartheta_S)=|\Delta_d|\cos(2\vartheta_S)\label{eq:13}
\\
Z_\Uparrow=[\varepsilon(1-W)+ \xi
(1+W)]^{\,2}+4W\,|\Delta(\vartheta_S)|^2\sin^2(\theta_{\,\alpha})\nonumber
\\
Z_\Downarrow
=[1+2W\cos(2\theta_{\alpha})+W^{\,2}]|\Delta(\vartheta_S)|^2
-4W\varepsilon^{2}\cos(2\theta_{\alpha})\nonumber
\\
-\,\frac{16 W^2\,(|\Delta(\vartheta_S)|^2-\varepsilon^2)
\varepsilon^{2}\sin^2(2\,\theta_{\,\alpha})}
{[1+2W\,\cos(2\,\theta_{\,\alpha})+W^{\,2}]\,|\Delta(\vartheta_S)|^2
-4W\varepsilon^{2}\cos(2\theta_{\alpha})}\nonumber
\\
 W\,=\,\sqrt{R_{\uparrow}R_{\downarrow}};\qquad\xi\,=\,\sqrt{\varepsilon^2
 \,-\,|\Delta(\vartheta_S)|^2}.\nonumber
\end{gather}
For $\gamma = \pi/4$:
\begin{gather}
\Delta(\vartheta_S)=|\Delta_d|\sin(2\vartheta_S)\label{eq:14}
\\
Z_\Uparrow=[\varepsilon(1+W)+ \xi
(1-W)]^{\,2}-4W\,|\Delta(\vartheta_S)|^2\sin^2(\theta_{\,\alpha})\nonumber
\\
Z_\Downarrow
=[1-2W\cos(2\theta_{\alpha})+W^{\,2}]|\Delta(\vartheta_S)|^2
+4W\varepsilon^{2}\cos(2\theta_{\alpha})\nonumber
\\
-\,\frac{16 W^2\,(|\Delta(\vartheta_S)|^2-\varepsilon^2)
\varepsilon^{2}\sin^2(2\,\theta_{\,\alpha})}
{[1-2W\,\cos(2\,\theta_{\,\alpha})+W^{\,2}]\,|\Delta(\vartheta_S)|^2
+4W\varepsilon^{2}\cos(2\theta_{\alpha})}.\nonumber
\end{gather}
For $\gamma = 0$, when $\theta_{\,\alpha}\,=\,0$, the expression for
the conductance obtained in Ref. \cite{Vod} follows from Eq.
(\ref{eq:12}). In the case of a nonmagnetic metal, when $ D_\uparrow
= D_\downarrow $ this expression is the same as that obtained in Ref.
\cite{Zaitsev}, and for $D\,=\,1/(1+Z^2)$ this expression is the same
as that obtained in Ref. \cite{Tinkham}. For $\gamma = \pi/4$, when
$\theta_{\,\alpha}\,=\,0$, the expression for the conductance
obtained in Ref. \cite{Vod2005} follows from Eq. (\ref{eq:12}).
\section{Andreev reflection}
The calculation of quasiclassical GFs in the expression for the
conductance allows one to conclude that for energies lower than
$|\Delta(\vartheta_S)|$ $(\varepsilon^{\,2}\,<\,|\Delta|^{\,2})$, the
following relation is true:
\begin{multline}\label{eq:15}
[1-\hat{g}_{\,s}^{\,A}\, \tau_z \,\hat{g}_{\,s}^{\,R}
\,\hat{\tau}_{\,z} -\hat{g}_{\,a}^{\,A}\,\hat{\tau}_z
\,\hat{g}_{\,a}^{\,R}\,\hat{\tau}_{\,z}
 +\hat{\Upsilon}_{\,s}^{\,A}\hat{\tau}_z
\hat{\Upsilon}_{\,s}^{\,R}\,\hat{\tau}_{\,z}\\
-\hat{\Upsilon}_{\,a}^{\,A} \hat{\tau}_z
\hat{\Upsilon}_{\,a}^{\,R}
\hat{\tau}_{\,z}]=4[-\hat{\tilde{g}}_{\,a}^{\,A}\,\hat{\tau}_z
\,\hat{\tilde{g}}_{\,a}^{\,R}\,\hat{\tau}_{\,z}]\sim \hat{1}.
\end{multline}
The comparison of the form of under-gap conductances in
Eq.(\ref{eq:1}) and that of the corresponding Eq. (25) in Ref.
\cite{Tinkham} shows that the matrix elements
$(\hat{\tilde{g}}_{\,a}^{\,R})^F$ and
$(\hat{\tilde{g}}_{\,a}^{\,A})^F$ are the amplitudes of the
Andreev reflection probability
$a(\varepsilon,\,\theta_{\,\alpha})$ in FIS contacts. Let us take the matrix elements of
$(\hat{\tilde{g}}_{\,a}^{\,R})^F$ given by Eq. (\ref{eq:11}) as $a(\varepsilon,\,\theta_{\,\alpha})$:
\begin{equation}\label{eq:16}
 a(\gamma,\,\varepsilon,\,\theta_{\,\alpha})= \frac{\sqrt{D_{\uparrow}D_{\downarrow}}\,\,
 \Delta(\vartheta_S)}{Z(\gamma)},
\end{equation}
where
\begin{multline*}
Z(0)= \\(1-\sqrt{R_\uparrow R_\downarrow})[\varepsilon
\cos(\theta_{\alpha})-\sqrt{|\Delta(\vartheta_S)|^2-\varepsilon^2}
\sin(\theta_{\alpha})]
\\
+i\,(1+\sqrt{R_\uparrow R_\downarrow})
[\sqrt{|\Delta(\vartheta_S)|^2-\varepsilon^2}\cos(\theta_{\,\alpha})+
\varepsilon \, \sin(\theta_{\,\alpha})].
\\
Z(\pi/4)=\hspace{6.5 cm} \\(1+\sqrt{R_\uparrow
R_\downarrow})[\varepsilon
\cos(\theta_{\alpha})-\sqrt{|\Delta(\vartheta_S)|^2-\varepsilon^2}
\sin(\theta_{\alpha})]
\\
+i\,(1-\sqrt{R_\uparrow R_\downarrow})
[\sqrt{|\Delta(\vartheta_S)|^2-\varepsilon^2}\cos(\theta_{\,\alpha})+
\varepsilon \, \sin(\theta_{\,\alpha})].
\end{multline*}
The presence of the imaginary part in functions
$a(\gamma,\,\varepsilon,\,\theta_{\,\alpha})$  means that Andreev
reflection is accompanied by the phase shift. The Andreev
reflection probability $A_\alpha(\gamma,\varepsilon)$\quad
($A_\alpha(\gamma,\varepsilon)\,=
\,a(\gamma,\,\varepsilon,\,\theta_{\,\alpha})\,a^{\,*}(\gamma,\,\varepsilon,\,\theta_{\,\alpha})$)
is:
\begin{gather}
 A_\alpha(\gamma,\varepsilon)\,=\,
 \frac{D_{\uparrow}\,D_{\downarrow}\,|\Delta(\vartheta_S)|^{\,2}}
 {|Z(\gamma)|^2} \label{eq:17}\\
  |Z(0)|^2\,=\,[1-\sqrt{R_\uparrow R_\downarrow}]^{\,2}\,|\Delta(\vartheta_S)|^{\,2}\nonumber\\
+\,4\,\sqrt{R_\uparrow R_\downarrow}\,\,
 [\sqrt{|\Delta(\vartheta_S)|^{\,2}\,-\,\varepsilon^{\,2}}\,\cos(\,\theta_{\,\alpha})\,
+\,\varepsilon\,\,\sin(\,\theta_{\,\alpha})]^{\,2}\nonumber.\\
|Z(\pi/4)|^2\,=\,[1-\sqrt{R_\uparrow R_\downarrow}]^{\,2}\,|\Delta(\vartheta_S)|^{\,2}\nonumber\\
+\,4\,\sqrt{R_\uparrow R_\downarrow}\,\,
 [\sqrt{|\Delta(\vartheta_S)|^{\,2}\,-\,\varepsilon^{\,2}}\,\sin(\,\theta_{\,\alpha})\,
-\,\varepsilon\,\,\cos(\,\theta_{\,\alpha})]^{\,2}\nonumber.
\end{gather}
From this equation it follows that: (1) the spin-mixing angle $\Theta
$ used in Refs. \cite{Sauls1988, Fogelstrom} corresponds, in our
notations, to $\theta_{\,\alpha}$ (for S/F/S and N/F/S contacts
$\Theta
$\,=\,$\theta_{\,\uparrow}^{\,r}-\theta_{\,\downarrow}^{\,r}$\,=\,
$\theta_{\,\uparrow}^{\,d}-\theta_{\,\downarrow}^{\,d}$ \quad
\cite{Barash, Fogelstrom}); (2) for $\gamma =0$, when
$\theta_{\,\alpha}\,<\,0$ the Andreev reflection probability of the
electron excitation with the spin projection $\alpha$ is larger than
that of the hole excitation; when $\theta_{\,\alpha}\,>\,0,$ the
Andreev reflection probability of the hole excitation with the spin
projection $\alpha$ is larger than that of the electron excitation;
for $\gamma = \pi/4,$ the relation is reverse; (3) the Andreev
reflection probability has maxima at
$\varepsilon\,=\,\varepsilon_{\,\alpha}^{\rm{b}}(\gamma)$ (at values
of the energy of electron (hole) excitations corresponding to the
energy levels of spin-dependent Andreev surface bound states).

The energy of spin-dependent bound states is:
\begin{gather}\label{eq:18}
\epsilon_{\alpha}^{\rm{b}}(0)=\left\{\begin{split}
-\,|\Delta(\vartheta_S)|\cos(\theta_{\alpha}) \,\,\, \mbox{for}\,
\,\,(\pi/2)\geq \theta_{\alpha}\geq0,
\\
|\Delta(\vartheta_S)|\cos(\theta_{\alpha})\,\,
\mbox{for}\,\,-(\pi/2)\leq\theta_{\alpha}\leq0,
\end{split}\right.
\\
\epsilon_{\alpha}^{\rm{b}}(\frac{\pi}{4})=\left\{\begin{split}
|\Delta(\vartheta_S)||\sin(\theta_{\alpha})|\,\,
\mbox{for}\,\,\,(\pi/2)\geq \theta_{\alpha}\geq0 \quad\\
-|\Delta(\vartheta_S)||\sin(\theta_{\alpha})| \,\, \mbox{for}
\,\,-(\pi/2)\leq\theta_{\alpha}\leq0.
\end{split}\right.\nonumber
\end{gather}
\begin{figure}
\centering \epsfxsize=7.2cm \epsffile{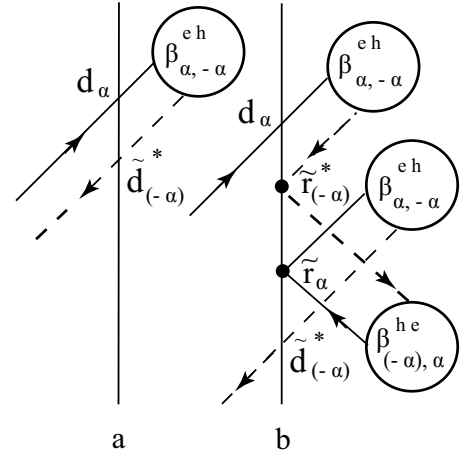} \caption{ Structure
of the diagrams corresponding to Andreev reflection in the
superconductor: diagram a),\, one-act process; diagram b),\, two-act
process. The vertex $\bigcirc$ is Andreev reflection of electron-like
(solid lines) and hole-like (broken lines) quasiparticles by the pair
patential. The vertex $\bullet$ is the normal reflection of
electron-like and hole-like quasiparticles by the barrier potential.
When the solid line transforms into the broken line,   $\bigcirc$
denotes the vertex $\beta_{\alpha,\,-\alpha}^{\,e\,h} $. When the
broken line transforms into the solid line, $\bigcirc$ denotes the
vertex $ \beta_{-\alpha,\alpha}^{\,h\,e}$. Parameters
$d_{\,\alpha},\,\tilde{d}_{\,\alpha},\,r_{\,\alpha}$ and
$\tilde{r}_{\,\alpha}$ are related as follows:
$\tilde{d}_{\,\alpha}\,=\,d_{\,\alpha}\,p_{\,x}^{\,S}/p_{\,x\,\alpha}^{\,F}$;
$\tilde{r}_{\,\alpha}\,=\,-\,r_{\,\alpha}^*\,d_{\,\alpha}/d_{\,\alpha}^*;\,\,
D_{\,\alpha}\,=\,d_{\,\alpha}\,\tilde{d}_{\,\alpha}^{\,*}$\,\,
\cite{Zaitsev}.}
\end{figure}
Spin-dependent Andreev surface bound states are formed in a
superconductor due to the interference of electron-like and hole-like
particles with different SDPS. One may demonstrate it by using a
phenomenological argument in Ref. \cite{Kashiwaya2}. Let us consider
diagrams in Fig. 1 corresponding to Andreev reflection of an electron
with the spin projection $\alpha$ and the energy less than $|\Delta|$
transmitted from a ferromagnet into a superconductor. The analysis of
these diagrams  and their summation  makes it possible to obtain the
following expression for a phenomenological expression of the
amplitudes of the Andreev reflection probability
$a(\varepsilon,\,\theta_{\,\alpha})$:
\begin{gather}
 a(\varepsilon,\,\theta_{\,\alpha})\,=\,
 d_{\,\alpha}\,\tilde{d}_{\,-\alpha}^*\,\beta_{\alpha,-\alpha}^{\,e\,h}[1\,+\,
 \tilde{r}_{\,-\alpha}^*\,\tilde{r}_{\,\alpha}\,
 \beta_{\alpha,\,-\alpha}^{\,e\,h}\,\beta_{-\alpha,\,\alpha}^{\,h\,e}\,\nonumber\\
+(\tilde{r}_{\,-\alpha}^*\,\tilde{r}_{\,\alpha}\,
\beta_{\alpha,\,-\alpha}^{\,e\,h}\beta_{-\alpha,\,\alpha}^{\,h\,e})^2+...]=
\,\frac{d_{\,\alpha}\,\tilde{d}_{\,-\alpha}^*\,\beta_{\alpha,\,-\alpha}^{\,e\,h}}
{1-\tilde{r}_{-\alpha}^*\tilde{r}_{\alpha}\,\beta_{\alpha,-\alpha}^{\,e\,h}\beta_{-\alpha,\alpha}^{\,h\,e}}\nonumber\\
=\frac{\sqrt{D_{\alpha}D_{-\alpha}\,p_{\,x\,\alpha}^{\,F}/p_{\,x\,-\,\alpha}^{\,F}}\,\,\,
e^{\,i\,\beta_{\,\alpha}^{\,r}}
\,\,\,\beta_{\alpha,\,-\alpha}^{\,e\,h}}
{e^{\,i\,\theta_{\,\alpha}}\,-\,
e^{\,-i\,\theta_{\,\alpha}}\sqrt{R_{\alpha}R_{-\alpha}}\,
\,\beta_{\alpha,-\alpha}^{\,e\,h}\,\beta_{-\alpha,\alpha}^{\,h\,e}}.\label{eq:19}
\end{gather}
The corresponding probability of Andreev reflection is:
\begin{gather}
    A(\varepsilon,\,\theta_{\,\alpha})\,=\,\frac{D_{\alpha}D_{-\alpha}\,p_{\,x,\,\alpha}^{\,F}/p_{\,x,\,-\,\alpha}^{\,F}
\,\,\beta_{\alpha,\,-\alpha}^{\,e\,h}\,\,\beta_{\alpha,\,-\alpha}^{*\,e\,h}}
{1+R_{\alpha}R_{-\alpha}\,\,|\beta_{\alpha,\,-\alpha}^{\,e\,h}|^{\,2}\,|\beta_{-\alpha,\alpha}^{\,h\,e}|^{\,2}-
Q} \nonumber
\\
Q=\sqrt{R_{\alpha}R_{-\alpha}}\left[\,\cos(2\,\theta_{\alpha})
[\beta_{\alpha,\,-\alpha}^{\,e\,h}\,\beta_{-\alpha,\,\alpha}^{\,h\,e}+
\beta_{\alpha,\,-\alpha}^{*\,e\,h}\,\beta_{-\alpha,\,\alpha}^{*\,h\,e}]\right.
\nonumber
\\
\left.+\,i\,\sin(2\,\theta_{\alpha}) [
\beta_{\alpha,\,-\alpha}^{*\,e\,h}\,\beta_{-\alpha,\,\alpha}^{*\,h\,e}-
\beta_{\alpha,\,-\alpha}^{\,e\,h}\,\beta_{-\alpha,\,\alpha}^{\,h\,e}]\right]\label{eq:20}
\end{gather}
By comparing formulas (\ref{eq:16},\,\ref{eq:17}), derived using
quasiclassical GFs, with formulas (\ref{eq:19},\,\ref{eq:20}),
obtained  using phenomenological arguments, we find the expressions
for the vertices $\beta_{\alpha,\,-\alpha}^{\,e\,h}$ and
$\beta_{-\alpha,\alpha}^{\,h\,e}.$ So for $\gamma = \pi/4$:
\begin{eqnarray}
   \beta_{\alpha,\,-\alpha}^{\,e\,h}=\sqrt{\frac{p_{\,x,\,-\,\alpha}^{\,F}}{p_{\,x,\,\alpha}^{\,F}}}\,\,
   \frac{\varepsilon-i\,
   \sqrt{|\Delta(\vartheta_S)|^{\,2}-
   \varepsilon^{\,2}}}{|\Delta(\vartheta_S)|}\,\frac{\Delta(\vartheta_S)}{|\Delta(\vartheta_S)|}\qquad \label{eq:21}
  \\
   \beta_{-\alpha,\alpha}^{\,h\,e}\,=\,-\,\sqrt{\frac{p_{\,x,\,\,\alpha}^{\,F}}{p_{\,x,\,-\,\alpha}^{\,F}}}\,\,
   \frac{\varepsilon\,-\,i\,\sqrt{|\Delta(\vartheta_S)|^{\,2}-\varepsilon^{\,2}}}
   {|\Delta(\vartheta_S)|}\,\frac{\,\,\Delta^*(\vartheta_S)}{|\Delta(\vartheta_S)|}\nonumber.
   \end{eqnarray}
For $\gamma = 0$ the expression for the vertex
$\beta_{-\alpha,\,\alpha}^{\,h\,e}$ is of an opposite sign. It
follows from formulas (\ref{eq:20}) and (\ref{eq:21}) that in the
absence of the interferential term $Q$ the probability of Andreev
reflection is a constant (independent of the energy $\varepsilon$)
quantity. The interference of electron-like and hole-like particles
reflected by the pair potential and the interface results in the
formation of spin-dependent Andreev surface bound states. For $\gamma
= 0$ at $\theta_{\,\alpha}\,=0\,$  the maximum in the probability of
Andreev reflection is at $\varepsilon\,=\,\pm\,|\Delta_d|$ as in
\cite{Tinkham}.  At $\theta_{\,\alpha}\,=\,\pm\,\pi/2\,$
spin-dependent Andreev surface bound states with the width $\Gamma$:
\begin{equation}\label{eq:22}
    \Gamma\,=\,\frac{(1-\sqrt{R{\uparrow}\,R{\downarrow}}\,)\,|\Delta(\vartheta_S)|}
{2\,\sqrt[4]{R{\uparrow}\,R{\downarrow}}}
\end{equation}
are formed at $\varepsilon\,=\,0$  on the Fermi level. For $\gamma =
\pi/4$ the spin degeneracy of the level on the Fermi surface
\cite{Hu} at  $\theta_{\,\alpha}\,\neq 0\,$ is removed. Two energy
levels symmetric with respect to the Fermi level are formed inside the
energy gap.
\section{Appearance of Andreev bound states in conductance of the  FIS contact}
We present below the results of numerical calculations of the charge
conductance of the FIS contact taking into account the phase shifts.
\begin{figure}
\centering \epsfxsize=8.0 cm \epsffile{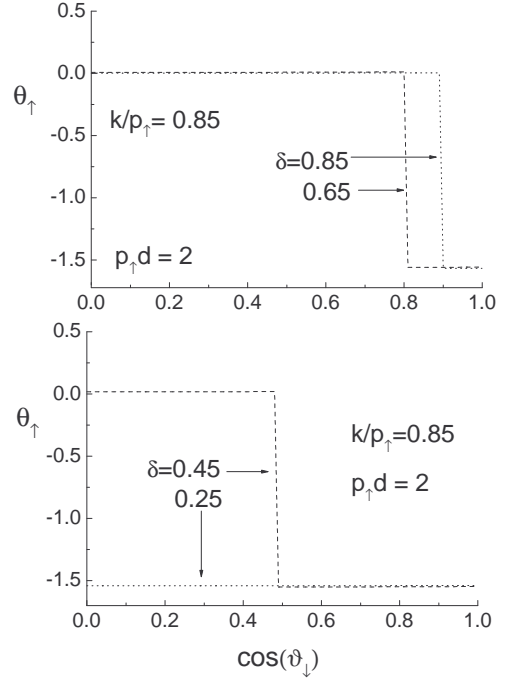}
\caption{Dependence of the angle  $\theta_{\,\uparrow}$  on
$\cos(\vartheta_{\,\downarrow})$ for various values of the
polarization of a ferromagnet $ \delta $ ($
\delta\,=\,p_\downarrow/p_\uparrow\,<1 $).}
\end{figure}
In the numerical calculations the relation between Fermi momenta of
contacting metals was the following:
$p_{\,S}\,=\,(p_{\,\uparrow}\,+\,p_{\,\downarrow})/2$. Calculations
are carried out for a rectangular barrier with a height $U$ counted
from the bottom of the conduction band of a superconductor. The
electron wave function in the isolator $\chi(x)$ is as follows:
$$
\chi(x)\,=\,C_1\,\exp(\,\mu_{\,x}
\,x)\,+\,C_2\,\exp(-\,\mu_{\,x}\,x),
$$
where
$\mu_{\,x}\,=\,\sqrt{k^2\,+\,p_{\,\parallel}^{\,2}};\,k^2\,=\,2m_b(U\,-\,E_F^S)\,$;
$E_F^S$   is the Fermi energy of a superconductor, $m_b$ is the
mass of an electron in a barrier.   In this case the expressions
for $ \theta_{\,\alpha}^{\,d}$ and $ \theta_{\,\alpha}^{\,r}$ have
the following form:
\begin{gather}
 \theta_{\alpha}^{\,d}=\widetilde{\theta}_{\,\alpha}^{\,d}-
 i\frac{1}{2}(p_{\,x,\,\alpha}^{\,F}\,+\,p_{\,x}^{\,S})\,d;
 \quad \theta_{\,\alpha}^{\,r}\,=\,\widetilde{\theta}_{\,\alpha}^{\,r}\,-\,ip_{\,x,\,\alpha}^{\,F}\,d\nonumber\\
 \widetilde{\theta}_{\,\alpha}^{\,d}\,=\,\arctan\left(\frac{(p_{\,x,\,\alpha}^{\,F}\,p_{\,\,x}^{\,S}-
 \mu_{\,x}^{\,2})\tanh(\mu_{\,x}\, d)}
 {\mu_{\,x}\,(p_{\,x,\,\alpha}^{\,F}\,+\,p_{\,x}^{\,S})}
  \right)\label{eq:23}\\
    \widetilde{\theta}_{\,\alpha}^{\,r}\,=\,\arctan \left(\frac{2\,
    \mu_{\,x}\,
    p_{\,x,\,\alpha}^{\,F}\,[\mu_{\,x}^{\,2}+(p_{\,x}^{\,S})^{\,2}]
    \tanh(\mu_{\,x}\,d)}{Z_{\,\alpha}} \right)\nonumber \\
    Z_{\,\alpha}=\mu_{x}^{\,2}\,[(p_{x}^{\,S})^{\,2}-(p_{x,\,\alpha}^{\,F})^{\,2}] \nonumber\\+
    [\mu_{x}^{\,4}-(p_{x}^{\,S}\,p_{x,\,\alpha}^{\,F})^{\,2}]\tanh^{2}(\mu_{x}
    d),\nonumber
    \end{gather}
so that the angle \,\,$\theta_{\,\alpha}$\,\,\,\,
$[\theta_{\,\alpha}=(\theta_{\alpha}^{\,r}-
\theta_{-\alpha}^{\,r})/2-
(\theta_{\,\alpha}^{\,d}-\theta_{-\,\alpha}^{\,d})]\,=\,(\widetilde{\theta}_{\alpha}^{\,r}-
\widetilde{\theta}_{-\alpha}^{\,r})/2-
(\widetilde{\theta}_{\,\alpha}^{\,d}-\widetilde{\theta}_{-\,\alpha}^{\,d})$
does not depend on the location of the barrier.

Figure 2 shows dependences of the angle $\theta_{\,\uparrow}$ on
$\cos(\,\vartheta_{\,\downarrow})$. All angles are connected by a
specular reflection $ p_\parallel $ =
$p_{\downarrow}\sin\vartheta_\downarrow $ =
$p_{\uparrow}\sin\vartheta_\uparrow = p_S\sin\vartheta_S $. Figure 2
shows that the angle $\theta_{\,\uparrow},$ being a combination of
phase shifts $ \theta_{\,\alpha}^{\,d}$ and
$\theta_{\,\alpha}^{\,r},$ has a jump for a part of electron
trajectories transmitted through the contact region. The jump in the
dependence of the angle $\theta_{\,\uparrow}$ on
$\cos(\,\vartheta_{\,\downarrow})$ is due to the jump in the
dependence of the phase shift $\widetilde{\theta}_{\,\alpha}^{\,r}$
on $\cos(\,\vartheta_{\,\downarrow})$.

By setting
$p_{\,x,\,\uparrow}^{\,F}=p_{\,x,\,\downarrow}^{\,F}=p_{\,x}^{\,S}=p_{\,x}^{\,N}$
in Eqs.(\ref{eq:23}), we get the following expression for phase shifts
$\widetilde{\theta}^{\,r}$ and $\widetilde{\theta}^{\,d}$, which correspond to the
 N/I/S contact with the same Fermi momenta
$p^{\,N}=p^{\,S}$  in normal metal $ (p^{\,N})$  and superconductor $(p^{\,S})$
$(p_{\,x}^{\,N}=p^{\,N}\cos(\,\vartheta ))$:
\begin{gather}
\widetilde{\theta}^{\,r}\,=\,\arctan \frac{2\,\mu_{\,x}
\,p_{\,x}^{\,N}}{[\mu_{\,x}^{\,2}-(p_{\,x}^{\,N})^{\,2}]\tanh(\mu_{\,x}\,d)}\label{eq:24}\\
\widetilde{\theta}^{\,d}\,=\,\arctan
\frac{[\mu_{\,x}^{\,2}-(p_{\,x}^{\,N})^{\,2}]\tanh(\mu_{\,x}\,d)}{2\,\mu_{\,x}
\,p_{\,x}^{\,N}};\,\,\widetilde{\theta}^{\,d}=\widetilde{\theta}^{\,r}+\frac{\pi}{2}.\nonumber
\end{gather}
One may see from this equation that the phase shift
$\widetilde{\theta}^{\,d}$ is continuous for the trajectory with
$p_{\,x}^{\,N}=\mu_{\,x}$, and the phase shift
$\widetilde{\theta}^{\,r}$ differs by $\pi$ for electron trajectories
with $p_{\,x}^{\,N}$, being somewhat larger $\mu_{\,x}$ and somewhat
smaller $\mu_{\,x}$. For the N/I/S contact, phase shifts
$\widetilde{\theta}_{\uparrow}^{\,r}$ and
$\widetilde{\theta}_{\downarrow}^{\,r}$ have jumps for the same
trajectory $\mu_{\,x}$. As a result, the angle
$\widetilde{\theta}_{\,\alpha}=0$. For the F/I/S contact, the
trajectories, at which  phase shifts
$\widetilde{\theta}_{\uparrow}^{\,r}$ and
$\widetilde{\theta}_{\downarrow}^{\,r}$ have jumps, are different.
These critical trajectories are the solutions of the transcendent
equation $Z_{\,\alpha}\,=\,0$ (see Eq. (23)). An analysis of the
numerical solution of this equation allows us to state that if
$kd\,\leq\,2$ and $k/p_{\uparrow}\,\leq\,1 $ there is always a set of
trajectories, for which the phase shift
$\widetilde{\theta}_{\uparrow}^{\,r}$ has a jump and the phase shift
$\widetilde{\theta}_{\downarrow}^{\,r}$ has no jump, or vice versa.
Thus, the angle $\widetilde{\theta}_{\,\alpha}$ has a jump being
equal to $\pm\,\pi/2$. Figure 2 shows the case when the phase shift
$\widetilde{\theta}_{\uparrow}^{\,r}$ has a jump and the phase shift
$\widetilde{\theta}_{\downarrow}^{\,r}$ has no jump.

As the polarization of a ferromagnet $\delta$
($\delta\,=\,p_\downarrow/p_\uparrow\,<1$) increases, the part of
electron trajectories with phase shifts experiencing a jump increases
as well, and at high values of the polarization of the ferromagnet
for all electron trajectories $\theta_{\,\uparrow}\sim\,-\pi/2,
\theta_{\,\downarrow}\sim\,+\pi/2$. For a rectangular model of the
potential barrier, the angle $\theta_{\,\alpha}$ is of an order of
$(\mp\pi/2)$  only for $kd\,\leq\,2$ and $k/p_{\uparrow}\,\leq\,1 $.
At $k/p_{\,\uparrow}\,>\,1$, the angle $\theta_{\,\alpha}<<1$.
\begin{figure}
\centering \epsfxsize=7.2cm \epsffile{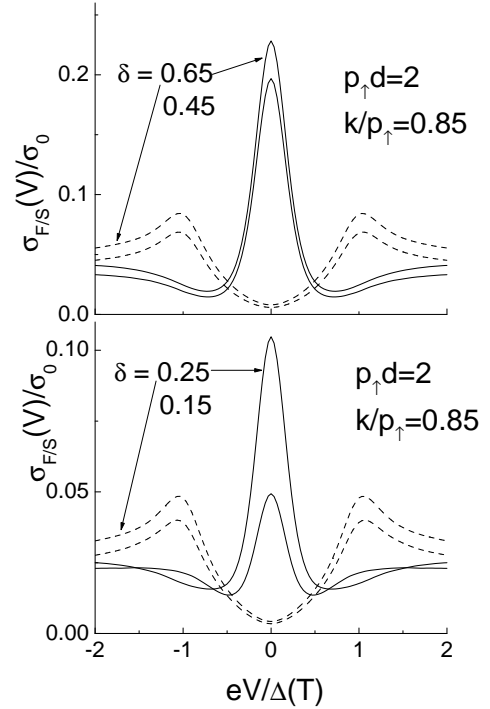}
\caption{Dependence of the normalized conductance
$\sigma_{\,F/S}(V)/\sigma_0$ ($\sigma_0
\,=\,e^{\,2}\,p_{\,\uparrow}/\,\pi^{\,2} $) from Eqs.(\ref{eq:12})
and (\ref{eq:13}) as a function of the applied potential for the $\{1
0 0\}$-oriented d - wave superconductor $(\gamma = 0)$ for various
values of the polarization of a ferromagnet $\delta$ not taking into
account (dashed lines) and taking into account (solid lines) the phase
shifts.}
\end{figure}
\begin{figure}
\centering \epsfxsize=7.8cm \epsffile{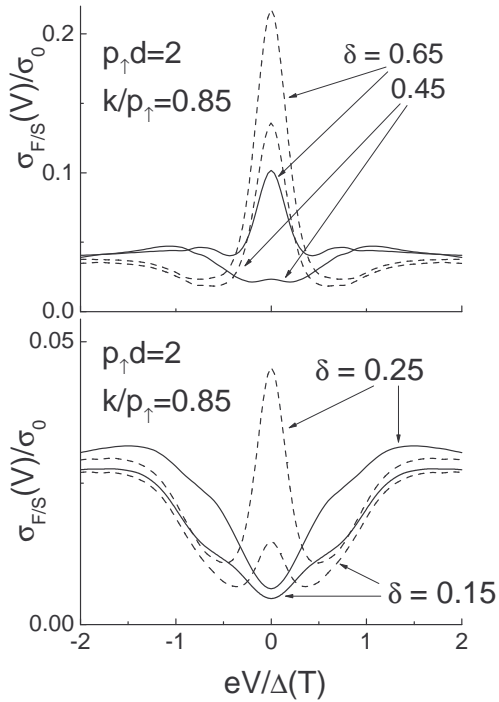}
\caption{Dependence of the normalized conductance
$\sigma_{\,F/S}(V)/\sigma_0$  from Eqs.(\ref{eq:12}) and
(\ref{eq:14})  as a function of the applied potential for the $\{1 1
0\}$-oriented d - wave superconductor $(\gamma = \pi/4)$ for
various values of the polarization of a ferromagnet $\delta$ not
taking into account (dashed lines) and taking into account (solid
lines) the angle $\theta_{\,\uparrow}$.}
\end{figure}

Figure 3 shows the results of numerical calculations of the
normalized conductance of the FIS contact
$\sigma_{\,F/S}(V)/\sigma_0$ for the  $\{1 0 0\}$ - oriented d - wave
superconductor taking into account and not taking into account the
phase shifts. Not taking into account the angle $\theta_{\,\alpha}$
(dashed lines), the plots illustrate the suppression of Andreev
reflection due to a decrease of the number of conducting channels
determined by the number of conducting channels in the subband with a
lower value of the Fermi momentum (in this case it is
$p_{\,\downarrow}$) as the polarization of the ferromagnet increases.
The appearance of electron trajectories with a jump of the angle
$\theta_{\,\alpha}$ forms Andreev bound states on the Fermi surface
($\varepsilon_\alpha^{\rm{b}}(0)=0$) (\ref{eq:18}). It results in the
appearance of a peak at a zero potential in the dependence
$\sigma_{\,F/S}(V)$. As $\delta$ increases, the part of electron
trajectories around the normal to the contact plane participating in
the formation of levels close to the Fermi level of a superconductor
increases (Fig. 1). However, it does not compensate the decrease of
the conductance at a zero potential due to the decrease of the number
of conducting channels. As a result, with increasing polarization of
the ferromagnet the zero-potential peak in the dependence
$\sigma_{\,F/S}(V)/\sigma_0$ decreases.

Figure 4 shows the results of numerical calculations of the
normalized conductance of the FIS contact
$\sigma_{\,F/S}(V)/\sigma_0$  for the $\{1 1 0\}$ oriented d - wave
superconductor $(\gamma = \pi/4)$. A part of electron trajectories
without a jump of the angle $\theta_{\,\alpha}$ forms the Andreev
bound state on the Fermi surface manifested in the zero-potential
conductance peak. The other part of electron trajectories with a jump
of the angle $\theta_{\,\alpha}$ forms the Andreev surface bound
state  with an energy of about $|\Delta_d\,\sin(2\vartheta_S)|$
manifested in the conductance peak at the potential close to
$|\Delta_d|$. At increasing polarization of the ferromagnet all
electron trajectories have a jump of the angle $\theta_{\,\alpha}.$
As a result, the conductance peak  at the zero potential disappears
and that at the potential close to $\pm
|\Delta_d\,\sin(2\vartheta_S)|$ increases. Plots in Fig. 4
demonstrate a tendency of the conductance peak to the decrease  at
the zero potential (the decrease of the part of electron trajectories
forming the level on the Fermi surface) and the increase of the
conductance at the potential close to the edge of the superconducting
gap (the increase of the part of electron trajectories forming the
jump of the phase shift $\theta_{\,\alpha}$) with increasing
polarization of the ferromagnet.
\section{Conclusion}
In this paper,  the influence of SDPS associated to the electronic
reflection and transmission amplitudes acquired by electrons upon
scattering on the potential barrier on the Andreev reflection
probability of electron and hole excitations for a
ferromagnet/isolator/d-wave superconductor contact and the charge
conductance of the ferromagnet/isolator/d-wave superconductor contact
as a function of the applied potential have been studied.
Spin-dependent Andreev bound states in a superconductor are found. It
is found that for parameters of a potential barrier $kd\,\leq\,2$ and
$k/p_{\uparrow}\,\leq\,1 $ there are always two groups of
trajectories of electron-like and hole-like quasiparticles, such that
due to the interference of one group of trajectories of electron-like
and hole-like quasiparticles reflected by the pair potential and the
interface, spin-dependent bound states are formed near the Fermi
level, whereas due to the interference of the other group of
trajectories spin-dependent bound states are formed in the vicinity
of the edges of the energy gap. As a result, SDPS can suppress the
zero-potential peak and restore finite-potential peaks in the charge
conductance of the F/I/d-wave superconductor contact for the
$\{110\}$ orientation of the d-wave superconductor and, on the
contrary, can restore the zero-potential peak and suppress
finite-potential peaks for the $\{100\}$ orientation of the d-wave
superconductor. The fitting of Eq. (\ref{eq:12}) to the experimental
dependence of the charge conductance of the FIS contact on the
applied potential makes it possible to determine the polarization of
a ferromagnet.

\section{ Acknowledgments}
I am grateful to G.B. Teitel'baum  for discussing the results of
this work. The work is supported by the Russian Foundation for Basic
Research, grant № 06-02-17233.

\appendix*
\section{Determining quasiclassic $\rm{GFs}$ $\hat{g}$ and
$\hat{\Upsilon}$.  Deriving equation (2).}

Let us start with equations for equilibrium thermodynamic GFs in the
matrix form \cite{Larkin}, taking into account the spin splitting of
the conduction band:
\begin{eqnarray}\label{eq:A1}
\left(i\varepsilon_n
\tau_{z}+\frac{1}{2m}\frac{\partial^2}{\partial{{\bf
r}^2}}+\hat{\Delta}({\bf r})+\hat{\mu}
 -\hat{\Sigma} \right)\hat{G}(\varepsilon_n,{\bf r},\;{\bf
r'})\nonumber \\ = \delta({\bf r}-{\bf r'}).\hspace{5cm}
\end{eqnarray}
Here $\hat{\Sigma}$ is the self-energy part which includes the
scattering by non-magnetic impurities and phonons  \cite{Larkin} . An
explicit form of this term is not needed for deriving the
quasiclassic equations. $\hat{G}(\varepsilon_n,{\bf r},\;{\bf r'})$
is the matrix temperature GF:
\begin{equation*}
\hat{G}(\varepsilon_n,{\bf r},\;{\bf r'})=
  \begin{pmatrix}
   G_{\alpha\alpha}&F_{\alpha\;-\alpha} \\
    F_{-\alpha\alpha}^{+} & \tilde{G}_{-\alpha\;-\alpha},
  \end{pmatrix}
 ;\,\hat{\mu}=\,\frac{1}{2m}
  \begin{pmatrix}
   p_{\alpha }^{\,2} &  \\
    & p_{-\alpha} ^{\,2}
  \end{pmatrix}.
\end{equation*}
$\hat{\tau}_{z}$ is the Pauli matrix; $\varepsilon_n = (2n+1)\pi$T is
the Matsubara frequency,  $ \alpha $ is the spin index;
$\hat{\Delta}({\bf r})$ is the  order parameter (as defined below
equation (2)); $ p_{\alpha}$ is the Fermi momentum; $m$ is the
electron mass; ${\bf r} = (x,\bf R)$, ${\bf R} = (y,z)$; $x$-axis is
perpendicular to the contact plane.

Passing to coordinates $\tilde{\rho}$ and $\rho$
$(\tilde{\rho}=\rho-\rho',\,\, 2\rho=R+R')$ in Eq. (\ref{eq:A1}) and
performing Fourier representation with respect to the $\tilde{\rho} $
coordinate, the following equation for $\hat{G}(x,x') =
\hat{G}(x,x',\rho,p_\|,\varepsilon_n)$ \;($p_\|$ is the momentum in
the contact plane) is obtained:
\begin{eqnarray}
\left(i\varepsilon_n
\tau_{z}+\frac{1}{2m}\frac{\partial^2}{{\partial{x}}^2} +i\frac{\bf
v_\|}{2}\frac{\partial}{\partial{\bf \rho}} +\frac{\hat{p}_x^2}{2m}
+\hat{\Delta}-\hat{\Sigma}\right)\hat{G}(x,x')\nonumber
\\
=\delta(x-x'),\hspace{5cm}\label{eq:A2}
\end{eqnarray}
In Eq. (A2): ${\bf v}_\|={\bf p}_\|/m,$\,\, $ \hat{p}_x=
 [\hat{p}_\alpha^2\,-\,{p_\|}^2]^{\,1/2}$.

Then the Zaitsev representation generalized for the description of metals
with a spin-split conduction band is used for the function
$\hat{G}(x,x')$:
\begin{eqnarray}\label{eq:A3}
    \hat{G}(x,x')=\sum_{n,\,m=1}^2 \hat{A}_k(x)\,
    \hat{G}_{n,\,m}(x,x')\,\hat{A}_n^*(x'), \qquad\\
    \hat{A}_k(x)=e^{\,-\,i(-\,1)^k\,\hat{p}_{\,x}\,x},\quad \hat{p}_{\,x}=\begin{pmatrix}
   p_{x,\,\alpha }& 0 \\
   0 & p_{x,\,{-\alpha}}\nonumber
  \end{pmatrix}
\end{eqnarray}
Representation (\ref{eq:A3}) explicitly takes account for oscillating
terms present in the function $\hat{G}(x,x')$ and waves of the $
\rm{exp} [\pm i\,(p_\uparrow x+p_\downarrow x')]$ type, arising from
partial reflection of the first electron of the superconducting pair
from the interface \cite{Ivanov}. Functions $\hat{G}_{\,n,\,m}(x,x')$
change at distances of an order of the mean free path of electrons in
a metal. By substituting Eq. (\ref{eq:A3}) to Eq. (\ref{eq:A2}) and,
neglecting the second $x$-derivative, we obtain an equation for slow
changing functions $\hat{G}_{\,k,\,n}(x,x')$:
\begin{eqnarray}
\hat{A}_k(x)\left( i\varepsilon_n\tau_{z}-i(-1)^k
\hat{v}_{x}\frac{\partial}{\partial{x}} +i\frac{\bf
v_\|}{2}\frac{\partial}{\partial{\bf \rho}}
 +\hat{\Delta}(x)\right.\qquad \label{eq:A4}\\
 \left.-\hat{\Sigma} \right) \hat{G}_{kn}(x,x')\hat{A}_n(x')=\delta(x-x'),
 \qquad \hat{v}_{x}=\frac{\hat{p}_{x}}{m}.\nonumber
\end{eqnarray}
Analogously, an equation conjugate to (\ref{eq:A1}) gives:
\begin{eqnarray}
\hat{A}_k(x)\hat{G}_{kn}(x,x')\hat{A}_n(x')\left(
i\varepsilon_n\tau_{z}+i\frac{{\bf v}_\|}{2}\frac{\partial}{\partial{\bf \rho}}
+\hat{\Delta}(x')-\hat{\Sigma} \right) \nonumber\\
   +\,\hat{A}_k(x)\frac{\partial
   \hat{G}_{kn}(x,x')}{\partial{x'}}\hat{A}_n(x')i(-1)^n \hat{v}_{x}=\delta(x-x').\label{eq:A5}
   \hspace{5mm}
\end{eqnarray}
In Eqs. (A4) and (A5) let us pass to functions $\hat{g}_{\,0}\equiv
\hat{g}_{\,0}(x,x') \equiv
\hat{g}_{\,0}(x,x',\rho,p_\|,\varepsilon_n)$ and
$\hat{\Upsilon}_{\,0}\equiv\hat{\Upsilon}_{\,0}(x,x') \equiv
\hat{\Upsilon}_{\,0}(x,x',\rho,p_\|,\varepsilon_n)$, being continuous
at a point $x = x'$, by using formulas: \setlength\arraycolsep{0pt}
\begin{eqnarray}
\hat{g}_{\,0}=\left\{\begin{split}
  \hat{g}_{0}^{>} =
  2i\sqrt{\hat{v}_{x}}\hat{G}_{11}(x,x')
 \sqrt{\hat{v}_{x}}
  -{\rm {sign}}(x-x');\, \hat{p}_{x}>0 \\
  \hat{g}_{0}^{<} =
  2i\sqrt{\hat{v}_{x}}\hat{G}_{22}(x,x')
 \sqrt{\hat{v}_{x}}
  + {\rm {sign}}(x-x');\,  \hat{p}_{x}<0
\end{split}\right.\nonumber\\
\hat{\Upsilon}_{\,0}=\left\{\begin{split} \hat{\Upsilon}_{0}^{>}=
  2i\sqrt{\hat{v}_{x}}\hat{G}_{12}(x,x')
 \sqrt{\hat{v}_{x}}, \,\,\hat{p}_{x}>0\\
  \hat{\Upsilon}_{0}^{<}=
  2i\sqrt{\hat{v}_{x}}\hat{G}_{21}(x,x')
 \sqrt{\hat{v}_{x}}, \,\, \hat{p}_{x}<0.
\end{split}\right.\hspace{2cm}
\end{eqnarray}
Let us call the obtained equations as (A4') and (A5'), respectively.
By subtracting equations (A5') from equations (A4') when $n = k$ and
adding equations (A4') and (A5') when $ n \neq k $, one may get
equations for functions $\hat{g}_{\,0}(x,x')$ and
$\hat{\Upsilon}_{\,0}(x,x')$. In these equations we set $x=x'$.
Finally, the following equations are obtained:
\begin{gather}
{\rm{sign}}(\hat{p}_{\,x})\hat{B}(x)\frac{\partial
\hat{g_0}}{\partial{\,x}}\,\hat{B}^{*}(x) +\frac{{\bf v_\|}}{2}\,
\frac{\partial}{\partial{\bf\rho}}\hat{B}(x)[\hat{v}_{\,x}^{-1},\,\hat{g_0}(x)]_{+}
\hat{B}^*(x)\nonumber\\
+[\hat{K}_{\,0},\,\hat{B}(x)\hat{g}_0\hat{B}^*(x)]_- = 0, \\
{\rm {sign}}(\hat{p}_{x})\hat{B}(x)\frac{\partial
\hat{\Upsilon}_0}{\partial{x}}\hat{B}(x)+\frac{{\bf
v_\|}}{2} \frac{\partial}{\partial{\bf\rho}}\hat{B}(x)[\hat{v}_{x}^{-1},\,\hat{\Upsilon}_0]_{-}\,\hat{B}(x)\nonumber\\
+
[\hat{K}_{\,0},\hat{B}(x)\hat{\Upsilon}_0\hat{B}(x)]_+ = 0,\\
\hat{B}(x)=e^{\,i\,{\rm {sign}}(\hat{p}_{\,x})\hat{p}_{\,x}\,x},\,\,
\hat{K}_{\,0}=\,-\,i\hat{v}_{\,x}^{-\frac{1}{2}}(i\varepsilon_n
\hat{\tau}_z +
\hat{\Delta}-\hat{\Sigma})\hat{v}_{\,x}^{-\frac{1}{2}},
\nonumber \\
\qquad \shoveleft{[a,\,b]_\pm = ab \pm ba.} \hspace{2cm}
\end{gather}
Considering that in expression for $\hat{B}(x)$ the matrix
$\hat{p}_{\,x}$ can be written with the help of the Pauli matrix
$\hat{\tau}_{\,x}$ as a sum of two components proportional to the
unit matrix and Pauli matrix $\hat{\tau}_{\,z}$:
\begin{equation}\label{eq:A10}
  \hat{p}_{\,x}
  =(\hat{p}_{\,x}+\hat{\tau}_x\hat{p}_{\,x}\hat{\tau}_x)/2+(\hat{p}_{\,x}-\hat{\tau}_x\hat{p}_{\,x}\hat{\tau}_x)/2,
 \end{equation}
 and going in Eqs. (A7) and (A8) to functions $\hat{g}$
 ($\hat{g} \equiv \hat{g}(\epsilon_{\,n},p_{\|}, \rho, x)$) and
  $\hat{\Upsilon}$ ($\hat{\Upsilon} \equiv \hat{\Upsilon}(\epsilon_{\,n},p_{\|}, \rho, x)$) by formulas:
\begin{gather}
\hat{g}=e^{\,i\,{\rm
{sign}}(\hat{p}_{\,x})\,\widehat{\Omega}\,x}\hat{g}_{\,0}
 e^{\,-\,i\,{\rm{sign}}(\hat{p}_{\,x}) \,\widehat{\Omega}\,x}\nonumber\\
\hat{\Upsilon}=e^{\,i\,{\rm{sign}}(\hat{p}_{\,x})\,\widehat{\Omega}\,x}\hat{\Upsilon}_{\,0}
e^{\,i\,{\rm{sign}}(\hat{p}_{\,x})\,\widehat{\Omega}\,x}\nonumber\\
\widehat{\Omega}=\frac{\hat{p}_{\,x}-\hat{\tau}_x\hat{p}_{\,x}\hat{\tau}_x}{2}
\label{eq:A11}
\end{gather}
one obtains equations (\ref{eq:2}).

If quasiclassic GFs $\hat{g}$ and $\hat{\Upsilon}$ are independent of the $\rho$ coordinate,
the condition $\hat{g}^{\,2}\,=\,1$ is met.

\end{document}